# Method of characteristics and Solution of DGLAP evolution equation in leading order (LO) and next to leading order (NLO) at small-x

R BAISHYA[1]

Physics Department, J.N. College, Boko-781123, Assam, India

J K SARMA[2]

Physics Department, Tezpur University, Napaam-784028, Assam, India

**Abstract.** In this paper the singlet and non-singlet structure functions have been obtained by solving Dokshitzer, Gribove, Lipatov, Alterelli, Parisi (DGLAP) evolution equations in leading order (LO) and next to leading order (NLO) at the small x limit. Here we have used a Taylor Series expansion and then the method of characteristics to solve the evolution equations. We have also calculated t and x-evolutions of deuteron structure function and the results are compared with the New Muon Collaboration (NMC) data.

## 1. Introduction

Deep Inelastic Scattering (DIS) [1-4] process is one of the basic processes for investigating the structure of hadrons. It is well known, all information about the structure of hadrons participating in DIS comes from the hadronic structure functions. According to QCD, at small value of x and at large values of $Q^2$ hadrons consist predominately of gluons and sea quarks, where x and $Q^2$ are Bjorken scaling variable and four momentum transfer in a DIS process respectively. The Dokshitzer, Gribov, Lipatov, Alterelli, Parisi (DGLAP) evolution equations [5-8] give t [$= \ln(Q^2/\Lambda^2)$, $\Lambda$ is the QCD cut off parameter] and x evolutions of structure functions. Hence the solutions of DGLAP evolution equations give quark and gluon structure functions that produce ultimately proton, neutron and deuteron structure functions.

---

[1] E-mail:- rjitboko@yahoo.co.in
[2] E-mail:- jks@tezu.ernet.in

There are various methods for solving DGLAP evolution equations and one of the limitation of these solutions is that, as the evolution equations are partial differential equation (PDE), their ordinary solutions are not unique solution, but a range of solutions, of course the range is very narrow. On the other hand, this limitation can be overcome by using method of characteristics. The method of characteristics is an important technique for solving initial value problems of first order PDE. It turns out that if we change co-ordinates from (x, t) to suitable new co-ordinates (S, τ) then the PDE becomes an ordinary differential equation (ODE) [13-15]. Then we can solve ODE by standard method.

In this paper, we obtain a solution of DGLAP equations for singlet and non-singlet structure functions at small-x at LO and NLO by using the method of characteristics. The result is compared with NMC data [16] for deuteron structure function. Here the section 1 is the introductions, section 2 deals with the necessary theory and section 3 is the results and discussion.

## 2. Theory

The DGLAP evolution equations for singlet and non-singlet structure functions in LO and NLO in standard form are

$$\frac{\partial F_2^S}{\partial t} - \frac{\alpha_s(t)}{2\pi}\left[\frac{2}{3}\{3 + 4\ln(1-x)\}F_2^S(x,t) + I_1^S(x,t) + I_2^S(x,t)\right] = 0, \quad (1)$$

$$\frac{\partial F_2^{NS}}{\partial t} - \frac{\alpha_s(t)}{2\pi}\left[\frac{2}{3}\{3 + 4\ln(1-x)\}F_2^{NS}(x,t) + I_1^{NS}(x,t)\right] = 0, \quad (2)$$

$$\frac{\partial F_2^S}{\partial t} - \frac{\alpha_s(t)}{2\pi}\left[\frac{2}{3}\{3 + 4\ln(1-x)\}F_2^S(x,t) + I_1^S(x,t) + I_2^S(x,t)\right] - \left(\frac{\alpha_s(t)}{2\pi}\right)^2 I_3^S = 0, (3)$$

$$\frac{\partial F_2^{NS}}{\partial t} - \frac{A_f}{t}\left[\{3 + 4\ln(1-x)\}F_2^{NS}(x,t) + I_1^{NS}(x,t)\right] - \left(\frac{\alpha_s(t)}{2\pi}\right)^2 I_2^{NS} = 0. \quad (4)$$

where

$$I_1^S(x,t) = \frac{4}{3}\int_x^1 \frac{d\omega}{1-\omega}\left[(1+\omega^2)F_2^S\left(\frac{x}{\omega},t\right) - 2F_2^S(x,t)\right], \quad (5a)$$

$$I_2^S(x,t) = N_f \int_x^1 \{\omega^2 + (1-\omega)^2\} G\left(\frac{x}{\omega},t\right) d\omega, \quad (5b)$$

$$I_3^S = \left[(x-1)F_2^S(x,t)\int_0^1 f(\omega)d\omega + \int_x^1 f(\omega)F_2^S\left(\frac{x}{\omega},t\right)d\omega + \int_x^1 F_{qq}^S(\omega)F_2^S\left(\frac{x}{\omega},t\right)d\omega + \int_x^1 F_{qg}^S(\omega)G\left(\frac{x}{\omega},t\right)d\omega\right]$$

(5c)

$$I_1^{NS}(x,t) = \frac{4}{3}\int_x^1 \frac{d\omega}{1-\omega}\left[(1+\omega^2)F_2^{NS}\left(\frac{x}{\omega},t\right) - 2F_2^{NS}(x,t)\right], \tag{5d}$$

$$I_2^{NS} = \left[(x-1)F_2^{NS}(x,t)\int_0^1 f(\omega)d\omega + \int_x^1 f(\omega)F_2^{NS}\left(\frac{x}{\omega},t\right)d\omega\right]. \tag{5e}$$

where

$$f(\omega) = C_F^2[P_F(\omega) - P_A(\omega)] + \frac{1}{2}C_F C_A[P_G + P_A(\omega)] + C_F T_R N_f P_{N_f}(\omega),$$

$$F_{qq}^S(\omega) = 2C_F T_T N_f F_{qq}(\omega),$$

$$F_{qg}^S(\omega) = C_F T_T N_f F_{qg}^1(\omega) + C_G T_T N_f F_{qg}^2(\omega),$$

$$P_F(\omega) = -\frac{2(1+\omega^2)}{(1-\omega)}\ln(\omega)\ln(1-\omega) - \left(\frac{3}{1-\omega} + 2\omega\right)\ln\omega - \frac{1}{2}(1+\omega)\ln\omega + \frac{40}{3}(1-\omega),$$

$$P_G(\omega) = \frac{(1+\omega^2)}{(1-\omega)}\left(\ln^2(\omega) + \frac{11}{3}\ln(\omega) + \frac{67}{9} - \frac{\pi^2}{3}\right) + 2(1+\omega)\ln\omega + \frac{40}{3}(1-\omega),$$

$$P_{N_f}(\omega) = \frac{2}{3}\left[\frac{1+\omega^2}{1-\omega}\left(-\ln\omega - \frac{5}{3}\right) - 2(1-\omega)\right],$$

$$P_A(\omega) = \frac{2(1+\omega^2)}{(1+\omega)}\int_{\left(\frac{\omega}{1+\omega}\right)}^{\left(\frac{1}{1+\omega}\right)} \frac{dk}{k}\ln\left(\frac{1-k}{k}\right) + 2(1+\omega)\ln(\omega) + 4(1-\omega),$$

$$F_{qq}(\omega) = \frac{20}{9\omega} - 2 + 6\omega - \frac{56}{9}\omega^2 + \left(1 + 5\omega + \frac{8}{3}\omega^2\right)\ln(\omega) - (1+\omega)\ln^2(\omega),$$

$$F_{qg}^1(\omega) = 4 - 9\omega - (1-4\omega)\ln\omega - (1-2\omega)\ln^2(\omega) + 4\ln(1-\omega) + \left[2\ln^2\left(\frac{1-\omega}{\omega}\right) - 4\ln\left(\frac{1-\omega}{\omega}\right) - \frac{2}{3}\pi^2 + 10\right]P_{qg}(\omega)$$

$$F_{qg}^2(\omega) = \frac{182}{9} + \frac{14}{9}\omega + \frac{40}{9\omega} + \left(\frac{136}{3}\omega - \frac{38}{3}\right)\ln\omega - 4\ln(1-\omega) - (2+8\omega)\ln^2\omega$$

$$+\left[-\ln^2\omega + \frac{44}{3}\ln\omega - 2\ln^2(1-\omega) + 4\ln(1-\omega) + \frac{\pi^2}{3} - \frac{218}{3}\right]P_{qg}(\omega) + 2Pqg(-\omega)\int_{\left(\frac{\omega}{1+\omega}\right)}^{\left(\frac{1}{1+\omega}\right)}\frac{dz}{z}\ln\frac{1-z}{z}$$

with $Pqg(\omega) = \omega^2 + (1-\omega)^2$, $C_A = C_G = 3$, $C_F(\omega) = \frac{(N_f^2 - 1)}{2N_f}$, $T_f = \frac{1}{2}$,

$\alpha_s(t) = \dfrac{4\pi}{\beta_0 t}\left[1 - \dfrac{\beta_1 \ln t}{\beta_0^2 t}\right]$. $\beta_0 = 11 - \dfrac{2}{3}N_f$ and $\beta_1 = 102 - \dfrac{38}{3}N_f$ are the one loop (LO) and two loop (NLO) correction to the QCD $\beta$ - function and $N_f$ being the flavors number. We can neglect $\beta_1$ for LO.

Let us introduce the variable u = 1-ω and note that $\dfrac{x}{\omega} = \dfrac{x}{1-u} = x\sum_{k=0}^{\infty} u^k$.

Since x<ω<1, so 0<u<1-x and hence the series is convergent for |u|<1. Now

$$\dfrac{x}{\omega} = \dfrac{x}{1-u} = \left(x + \dfrac{xu}{1-u}\right).$$

So, using Taylor's expansion series we can rewrite $F_2^S\left(\dfrac{x}{\omega}, t\right)$ and $G\left(\dfrac{x}{\omega}, t\right)$ as

$$F_2^S\left(\dfrac{x}{\omega}, t\right) = F_2^S\left(x + \dfrac{xu}{1-u}, t\right)$$

$$= F_2^S(x,t) + \dfrac{xu}{1-u}\dfrac{\partial F_2^S(x,t)}{\partial x} + \dfrac{1}{2}\left(\dfrac{xu}{1-u}\right)^2 \dfrac{\partial^2 F_2^S(x,t)}{\partial x^2} + \ldots$$

Since x is small in our region of discussion, the terms containing $x^2$ and higher powers of x can be neglected. i.e.

$$F_2^S\left(\dfrac{x}{\omega}, t\right) = F_2^S(x,t) + \dfrac{xu}{1-u}\dfrac{\partial F_2^S(x,t)}{\partial x} \tag{6a}$$

$$G\left(\dfrac{x}{\omega}, t\right) = G(x,t) + \dfrac{xu}{1-u}\dfrac{\partial G(x,t)}{\partial x} \tag{6b}$$

Now let us assume G (x, t) = k(x) $F_2^S$ (x, t), where k (x) is a suitable function of x or may be a constant. We may assume k (x) = k, $ax^b$, $ce^{dx}$ where k, a, b, c, d are suitable parameters which can be determined by phenomenological analysis. Using equation (6a), (6b) in equation (5a) and (5b) and performing u-integrations, equation (1) becomes the form

$$-t\dfrac{\partial F_2^S(x,t)}{\partial t} + A_f L(x) F_2^S(x,t) + A_f M(x)\dfrac{\partial F_2^S}{\partial x} = 0 \tag{7}$$

where

$$A_f = \dfrac{\alpha_s(t)}{3\pi}t = \dfrac{4}{3\beta_0} = \dfrac{4}{33 - 2N_f}$$

L (x) = A (x) + k (x) C (x) + D (x) ∂ k(x)/ ∂ x, (8a)

M (x) = B (x) + k (x) D (x), (8b)

$A(x) = 2x + x^2 + 4\ln(1-x)$, (8c)

$$B(x) = x - x^3 - 2x \ln(x), \qquad (8d)$$

$$C(x) = 2N_f \left( \frac{2}{3} - x + x^2 + -\frac{2}{3}x^3 \right), \qquad (8e)$$

$$D(x) = 2N_f \left[ -\frac{5}{3}x + 3x^2 - 2x^3 + \frac{2}{3}x^4 - x \ln(x) \right]. \qquad (8f)$$

To introduce method of characteristics, let us consider two new variables S and $\tau$ instead of x and t, such that

$$\frac{dt}{dS} = -t, \qquad (9a)$$

$$\frac{dx}{dS} = A_f M(x) \qquad (9b)$$

which are known as characteristics equations. Putting these equations in (7), we get

$$\frac{dF_2^S(S, \tau)}{dS} + L(S, \tau) F_2^S(S, \tau) = 0. \qquad (10)$$

This can be solved as-

$$F_2^S(S, \tau) = F_2^S(\tau) \left( \frac{t}{t_0} \right)^{L(S, \tau)}, \qquad (11)$$

where $L(S, \tau) = A_f \cdot L(x)$ and $F_2^S(S, \tau) = F_2^S(\tau)$; S = 0, t = $t_0$.

Now we have to replace the co-ordinate system (S, $\tau$) to (x, t) in equation (11) with the input function $F_2^S(\tau) = F_2^S(x, t_0)$ and will get the t- evolution of singlet structure function in the LO as

$$F_2^S(x, t) = F_2^S(x, t_0) \left( \frac{t}{t_0} \right)^{A_f L(x)}. \qquad (12a)$$

Similarly the x- evolution of singlet structure function will be

$$F_2^S(x, t) = F_2^S(x_0, t) \exp\left[ -\int_{x_0}^{x} \frac{L(x)}{M(x)} dx \right]. \qquad (12b)$$

Proceeding in the same way, we get t and x evolutions of non-singlet structure function from equation (2) as

$$F_2^{NS}(x, t) = F_2^{NS}(x, t_0) \left( \frac{t}{t_0} \right)^{A(x)}, \qquad (13a)$$

$$F_2^{NS}(x,t) = F_2^{NS}(x_0,t)\exp\left[-\int_{x_0}^{x}\frac{A(x)}{B(x)}dx\right]. \tag{13b}$$

The deuteron, proton and neutron structure functions measured in DIS can be written in terms of singlet and non-singlet quark distribution functions as

$$F_2^d(x,t) = \tfrac{5}{2}F_2^S(x,t), \tag{14a}$$

$$F_2^p(x,t) = \frac{5}{18}F_2^S(x,t) + \frac{3}{18}F_2^{NS}(x,t), \tag{14b}$$

$$F_2^N(x,t) = \frac{5}{18}F_2^S(x,t) - \frac{3}{18}F_2^{NS}(x,t). \tag{14c}$$

The t and x-evolution of deuteron structure functions in LO can be obtained as

$$F_2^d(x,t) = F_2^d(x,t_0)\left(\frac{t}{t_0}\right)^{A_f L(x)}, \tag{15a}$$

$$F_2^d(x,t) = F_2^d(x_0,t)\exp\left[-\int_{x_0}^{x}\frac{L(x)}{M(x)}dx\right], \tag{15b}$$

where $F_2^d(x,t_0) = \frac{5}{2}F_2^S(x,t_0)$ and $F_2^d(x_0,t) = \frac{5}{2}F_2^S(x_0,t)$ are input functions.

Similarly in the NLO, the t and x evolution of deuteron structure functions will be obtain as

$$F_2^d(x,t) = F_2^d(x,t_0)\left(\frac{t}{t_0}\right)^{\frac{3}{2}A_f[L(x)+T_0 L_1(x)]}, \tag{16a}$$

$$F_2^d(x,t) = F_2^d(x_0,t)\exp\left[-\int_{x_0}^{x}\frac{L(x)+T_0 L_1(x)}{M(x)+T_0 M_1(x)}dx\right], \tag{16b}$$

where

$L_1(x) = B_1(x) + k(x) B_2(x) + B_4(x) \partial k(x)/\partial x,$ (17a)

$M_1(x) = B_3(x) + k(x) B_4(x),$ (17b)

$$B_1(x) = x\int_0^1 f(\omega)d\omega - \int_0^x f(\omega)d\omega + \frac{4}{3}N_f\int_x^1 F_{qq}(\omega)d\omega, \tag{17c}$$

$$B_2(x) = \int_x^1 F_{qg}^S(\omega)d\omega, \tag{17d}$$

$$B_3(x) = x\int_x^1\left[f(\omega) + \frac{4}{3}N_f F_{qg}^S(\omega)\right]\frac{1-\omega}{\omega}d\omega, \tag{17e}$$

$$B_4(x) = x \int_x^1 \frac{1-\omega}{\omega} F_{qg}^S(\omega) d\omega. \qquad (17f)$$

Here we consider an extra assumption $\left(\frac{\alpha_s(t)}{2\pi}\right)^2 = T_0 \left(\frac{\alpha_s(t)}{2\pi}\right)$, where $T_0$ is a numerical parameter. By a suitable choice of $T_0$ we can reduce the error to a minimum.

## 3. Results and Discussion

In this paper, we compare our result of t and x evolution of deuteron structure function $F_2^d$ measured by the NMC in muon deuteron DIS with incident momentum 90, 120, 200, 280 GeV [16]. For quantitative analysis, we consider the QCD cut-off parameter $\Lambda_{\overline{MS}} = 0.323$ GeV for $\alpha_S(M_Z^2) = 0.119 \pm 0.002$ and $N_f = 4$ [18]. It is observed that our result is very sensitive to arbitrary parameters k, a, b, c and d in t-evolution. In figure (2-4) for t-evolution, we have plotted computed values of $F_2^d$ against $Q^2$ values for a fixed x in LO and NLO. The best fitting ranges are $1.0 \leq k \leq 1.4$, $8.0 \leq a \leq 9.0$, $0.6 \leq b \leq 0.8$, $2.0 \leq c \leq 6.5$, $-28 \leq d \leq -10$ in LO and $0.5 \leq k \leq 0.7$, $1.2 \leq a \leq 6.0$, $0.0001 \leq b \leq 0.002$, $1.0 \leq c \leq 5.0$, $0.004 \leq d \leq 0.3$ in NLO. Here the solid lines represent the best fitting curves in NLO and the dotted lines represent those for LO evolutions.

In figure (5-7) for x-evolution, we have plotted computing values of $F_2^d$ against the x values for a fixed $Q^2$. Here we have plotted the graphs for $Q^2 = 11.5, 15, 20$ and 27 GeV$^2$ for the range of $0.07 \leq x \leq 0.18$. Here we have considered the input parameter $x_0 = 0.18$. The best fitting ranges are $0.5 \leq k \leq 0.7$, $1.5 \leq a \leq 2.2$, b=1.0, $0.3 \leq c \leq 0.4$, $1.0 \leq d \leq 2.5$ in LO and $2.0 \leq k \leq 3.0$, $2.0 \leq a \leq 2.8$, $0.01 \leq b \leq 0.08$, $2.0 \leq c \leq 2.5$, $0.05 \leq d \leq 3.0$ in NLO. Here also the solid lines represent the best-fit curves for NLO and the dotted lines represent for LO evolutions.

Though there are various methods to solve DGLAP evolution equation to calculate quark and gluon structure functions, our method of characteristics to solve these equations is also a viable alternative. Though mathematically vigorous, it changes the integro-differential equations into ODE and then makes it possible to obtain unique solutions.

**Figure captions**

**Figure 1(a).** Characteristic curves for constant values of $\tau$ ($\tau_1$, $\tau_2$, $\tau_3$ . . .). The values of S change along a vertical characteristic curve. On the other hand, along a horizontal characteristic curve, the values of $\tau$ change for constant values of S ($S_1$, $S_2$, $S_3$ . . .) in the t-x plane.

**Figure 1(b).** $T^2(t)$ and $T_0.T(t)$, where $T(t) = \left(\dfrac{\alpha_s(t)}{2\pi}\right)$, against $Q^2$ in the range $0 \leq k \leq 30$.

**Figure 2-4.** Results of t-evolution of deuteron structure function $F_2^d$ for the given value of x for k(x) = constant (k), k(x) = $ax^b$, and k(x) = $ce^{dx}$. Data points at lowest-$Q^2$ values in the figures are taken as input to test the evolution equations (15a) and (16a) for LO and NLO respectively.

**Figure 5-6.** Results of x-evolution of deuteron structure function $F_2^d$ for the given values of t for k(x) = constant (k), k(x) = $ax^b$, and k(x) = $ce^{dx}$. Data point at lowest-x values in the figures are taken as inputs to test the evolution equation (15b) and (16b).

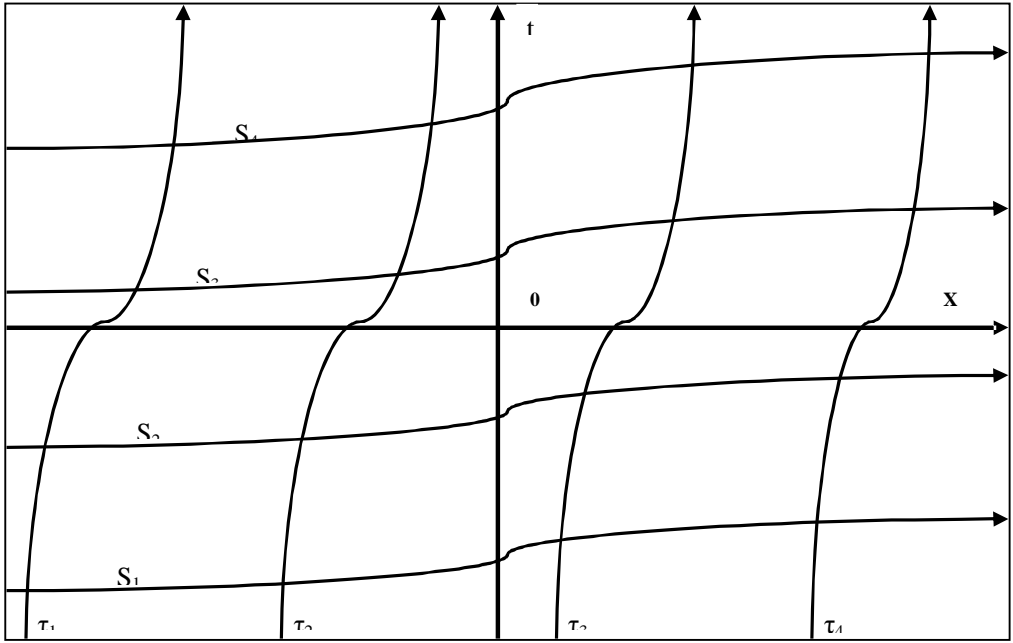

**Figure 1(a). Characteristics curves.**

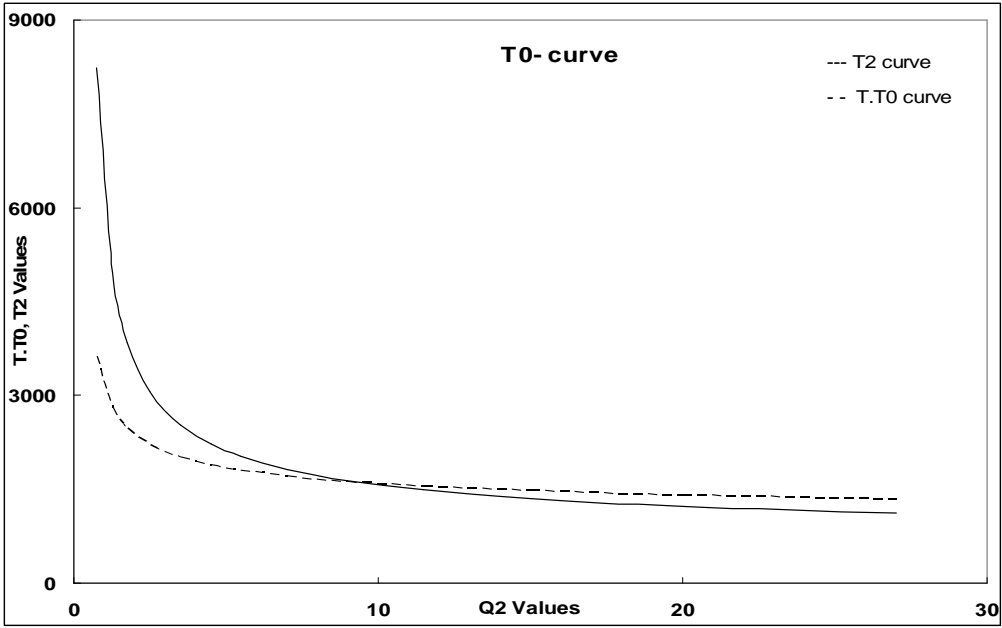

**Figure 1(b). $T^2$ and $T_0.T$ curves.**

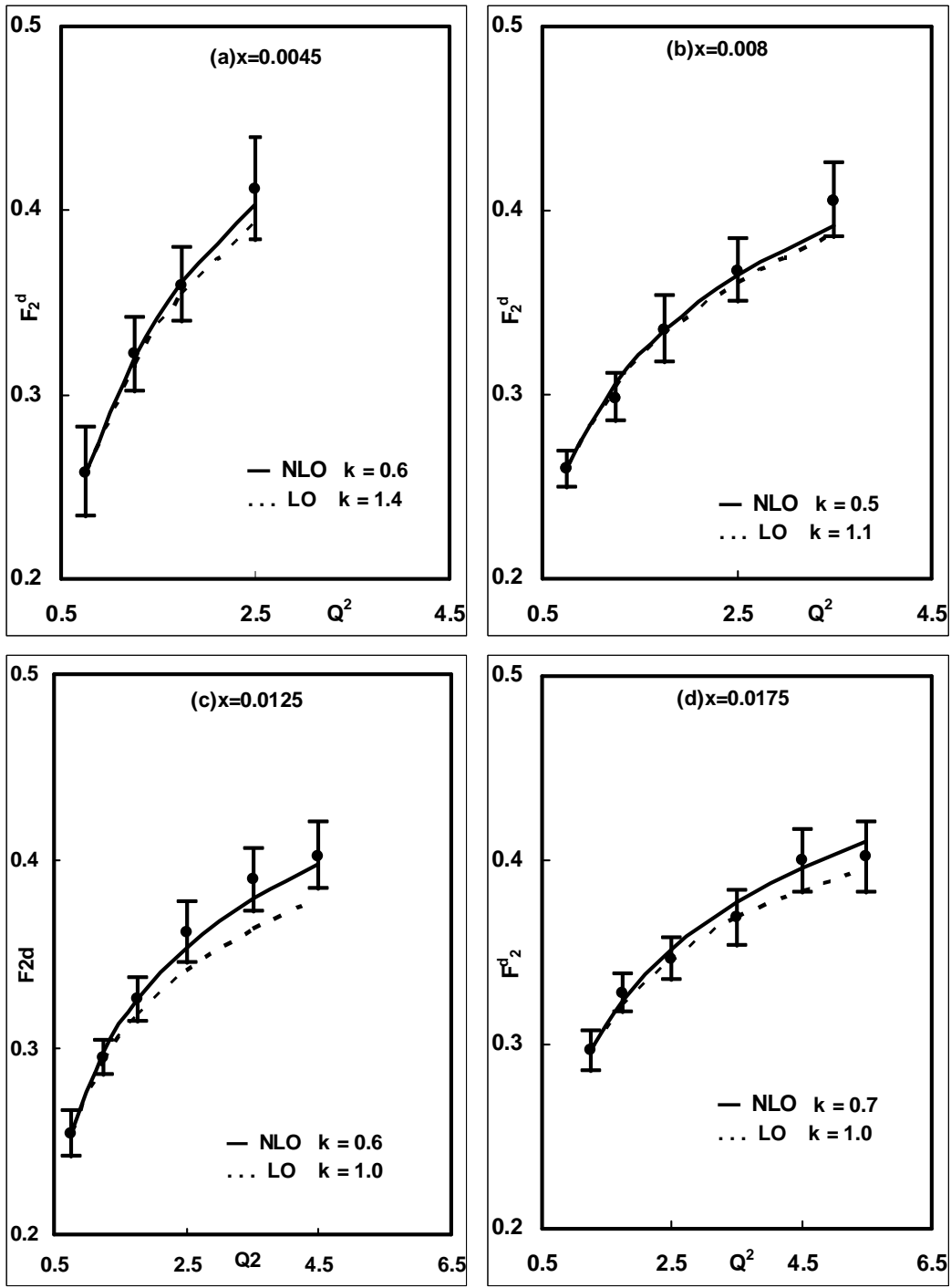

**Figure 2 (a-d): t-Evolution of deuteron structure function in LO and NLO as k is a constant.**

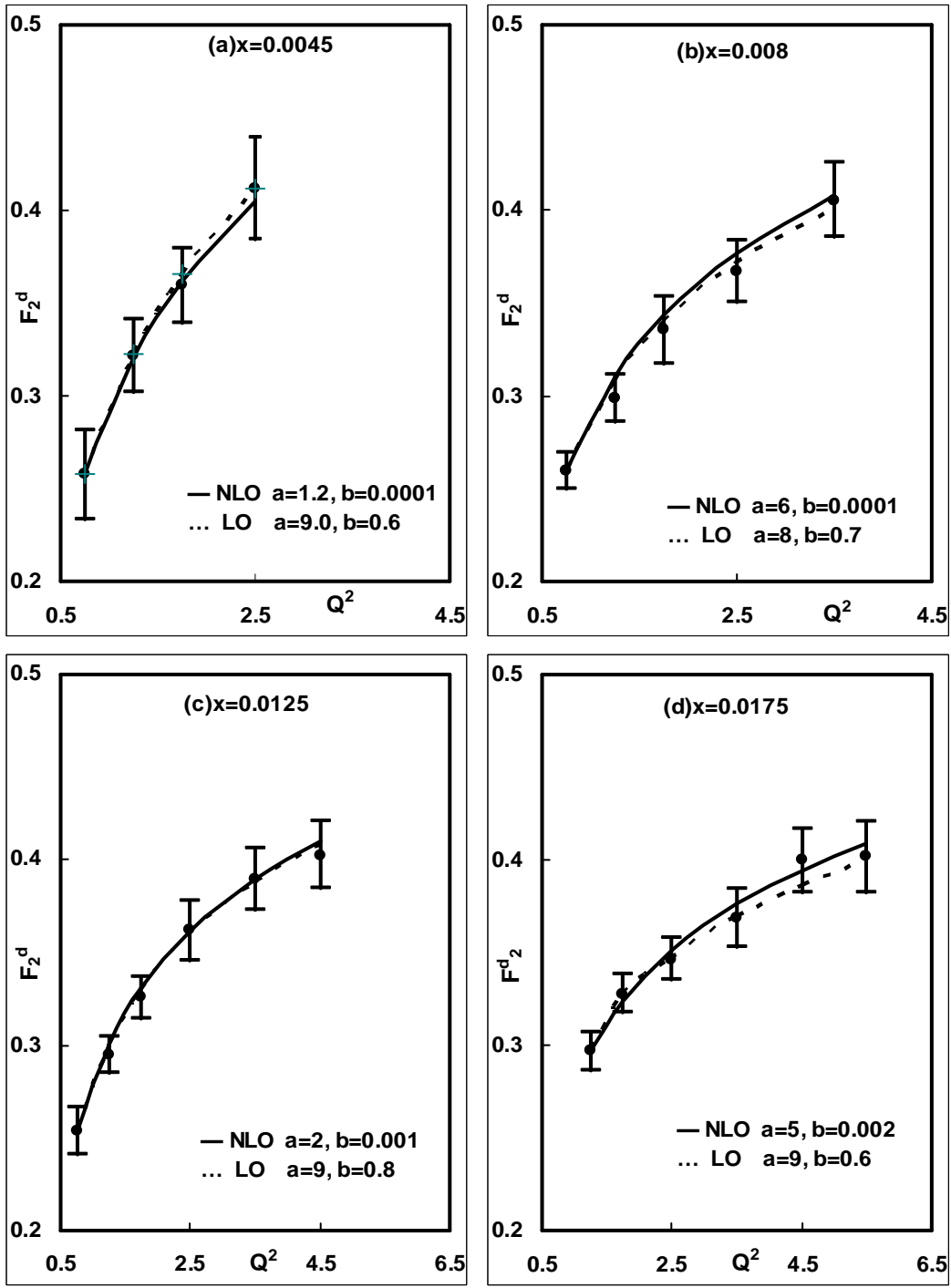

**Figure 3 (a-d): t-Evolution of deuteron structure function in LO and NLO as k is a power function of x.**

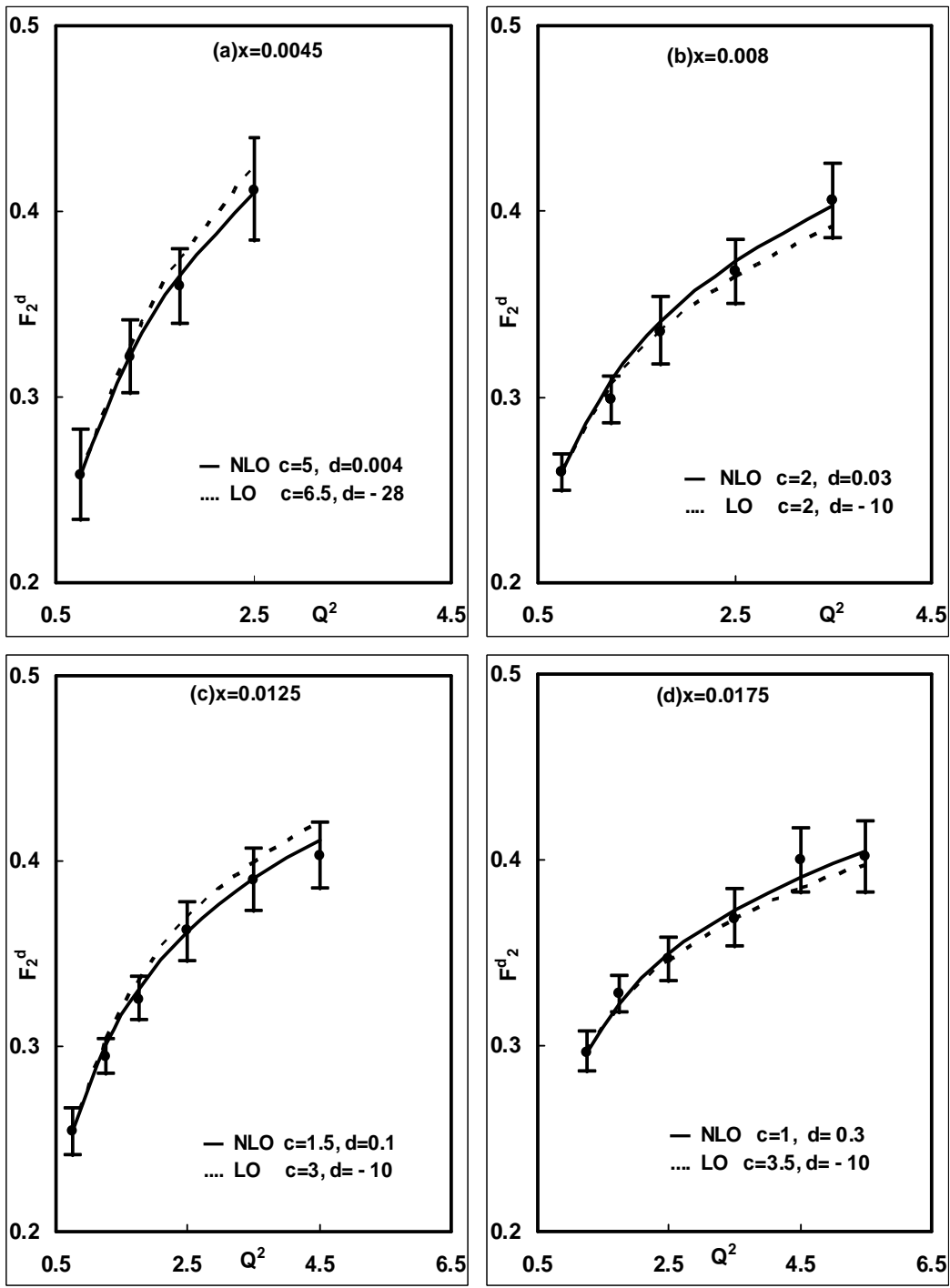

**Figure 4 (a-d): t-Evolution of deuteron structure function in LO and NLO as k is an exponential function of x.**

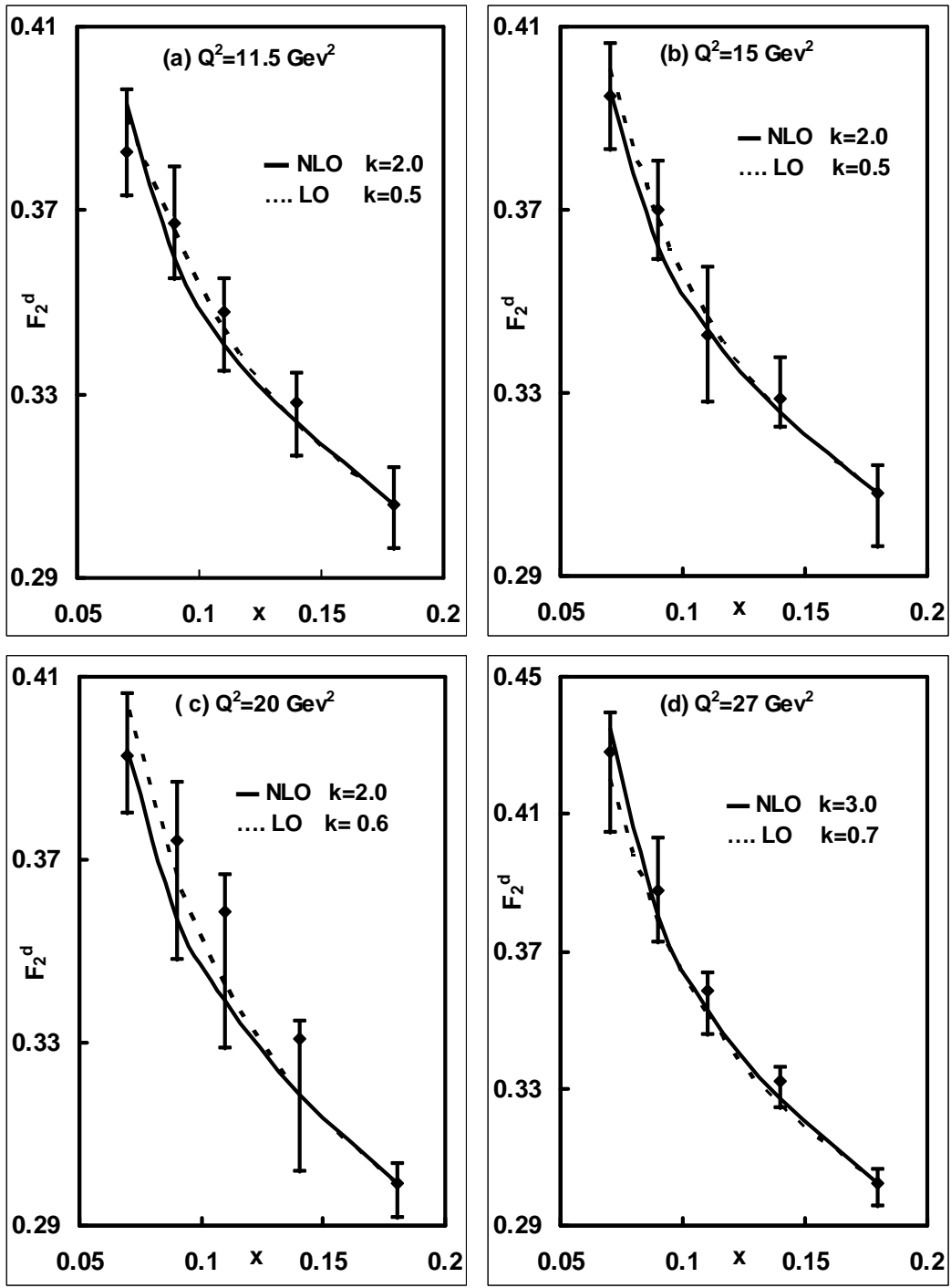

**Figure 5 (a-d): x-Evolution of deuteron structure function in LO and NLO as k is a constant.**

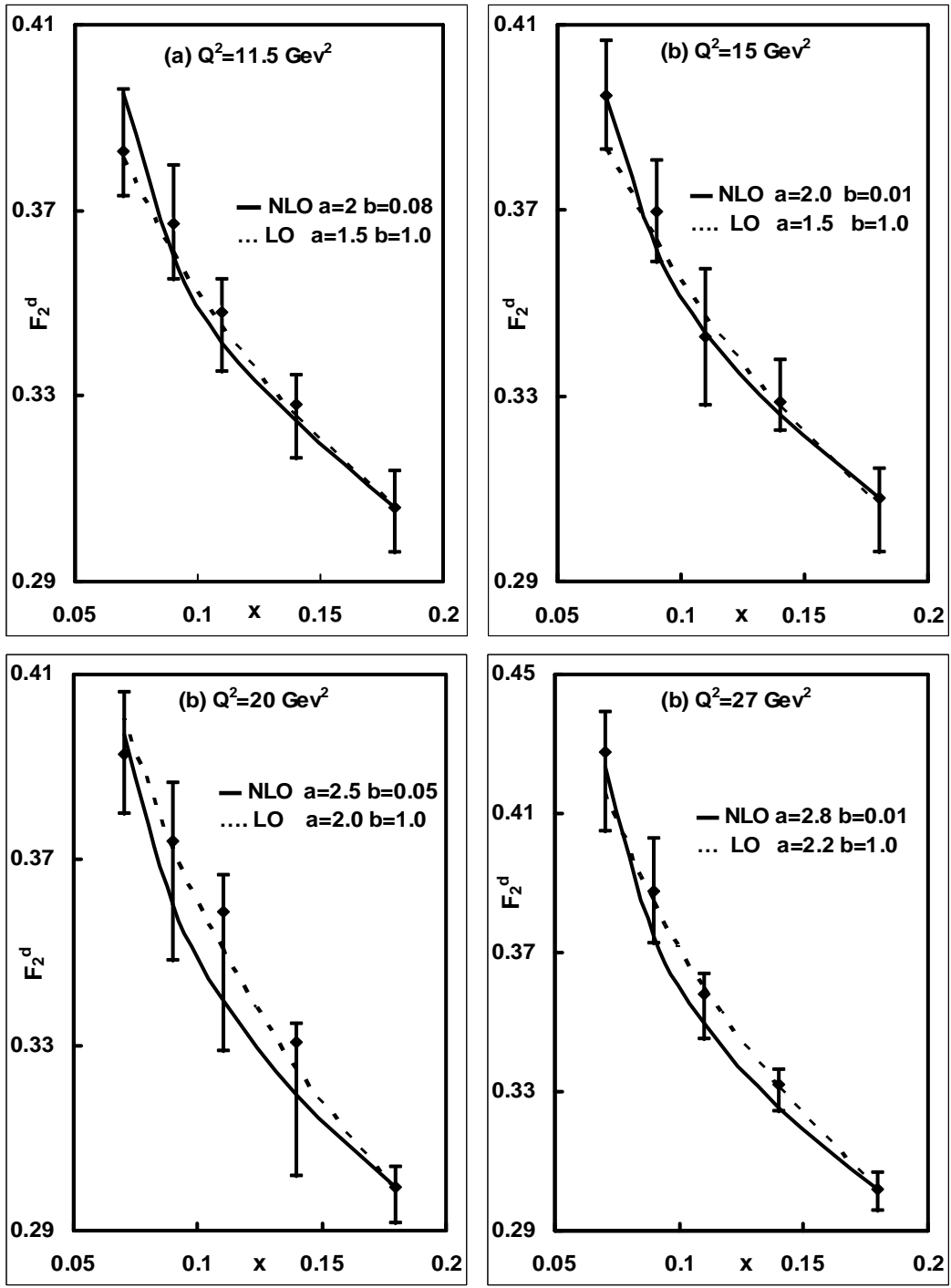

**Figure 6 (a-d): x-Evolution of deuteron structure function in LO and NLO as k is a power function of x.**

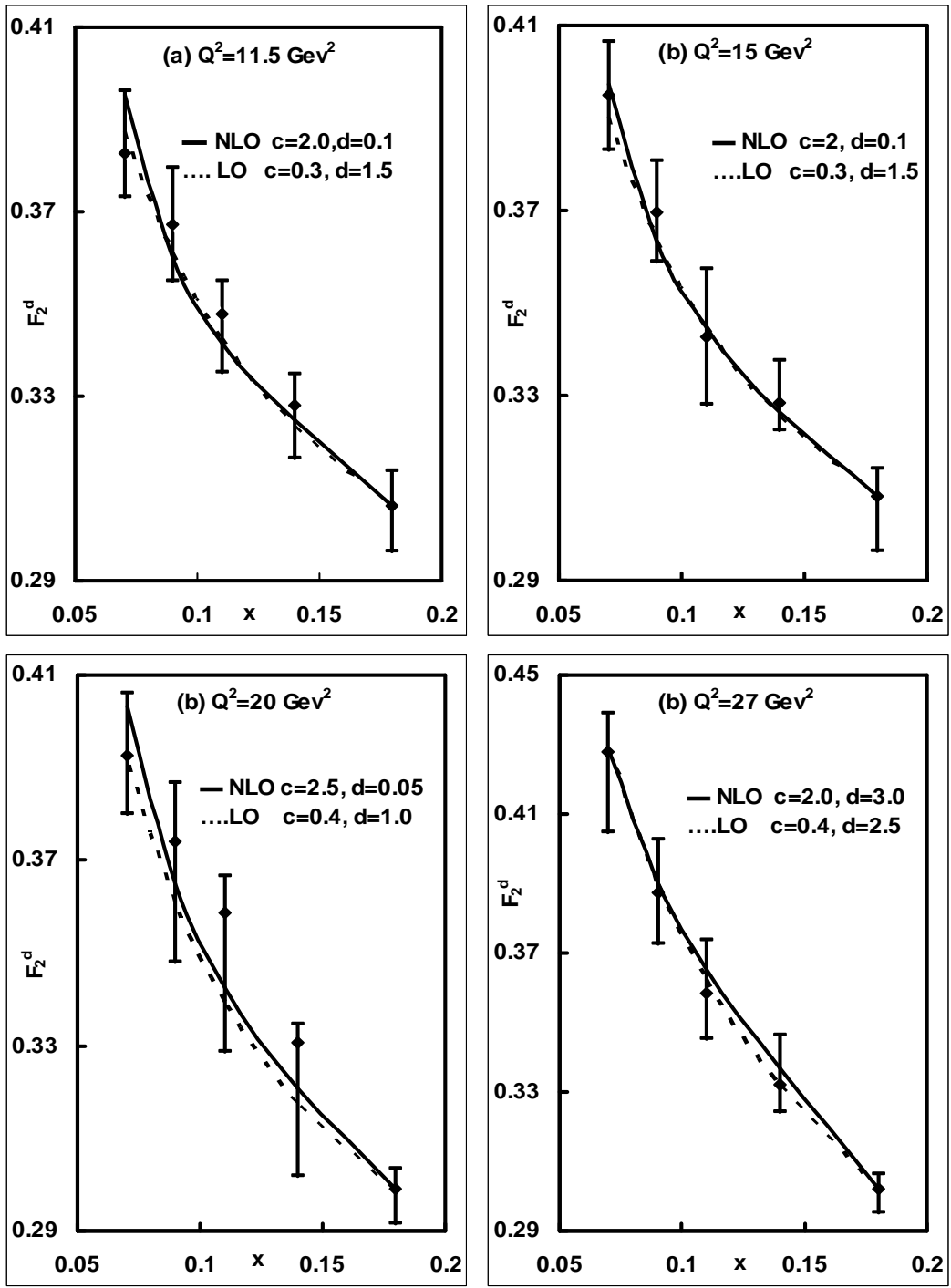

**Figure 7 (a-d): x-Evolution of deuteron structure function in LO and NLO as k an exponential function of x.**